\documentclass{aa520}

\usepackage{graphicx}
\usepackage{times}
\usepackage{dcolumn}
\usepackage{natbib}

\bibpunct{(}{)}{;}{a}{}{,}
\newcolumntype{.}{D{.}{.}{-1}}

\begin{document}


\title{Integral-field spectrophotometry of the quadruple QSO HE~0435$-$1223:
  Evidence for microlensing}

\author{L. Wisotzki\inst{1,2} 
        \and 
        T. Becker\inst{1}
        \and
        L. Christensen\inst{1}
        \and
        A. Helms\inst{2}
        \and
        K. Jahnke\inst{1}
        \and
        A. Kelz\inst{1}
        \and
        M. M. Roth\inst{1}
        \and
        S. F. Sanchez\inst{1}
        }

\authorrunning{L. Wisotzki et al.}
\titlerunning{Spectrophotometry of HE~0435$-$1223}

\institute{%
           Astrophysikalisches Institut Potsdam, An der Sternwarte 16,
           14482 Potsdam, Germany,
           email: lwisotzki@aip.de
           \and
           Universit\"at Potsdam, Am Neuen Palais 10, 14469 Potsdam, Germany, 
          }

\date{Revised \today}

\abstract{We present the first spatially resolved spectroscopic
observations of the recently discovered quadruple QSO and gravitational
lens HE~0435$-$1223. Using the Potsdam Multi-Aperture Spectrophotometer (PMAS),
we show that all four QSO components have very similar but not identical spectra.
In particular, the spectral slopes of components A, B, and D are indistinguishable, 
implying that extinction due to dust plays no major role in the lensing galaxy.
While also the emission line profiles are identical within the error bars,
as expected from lensing, the equivalent widths show significant
differences between components. Most likely, microlensing is responsible
for this phenomenon. This is also consistent with the fact that component D,
which shows the highest relative continuum level, 
has brightened by 0.07~mag since Dec~2001.
We find that the emission line flux ratios between
the components are in better agreement with simple lens models than
broad band or continuum measurements, but that the discrepancies between
model and data are still unacceptably large.
Finally, we present a detection of the lensing galaxy, 
although this is close to the limits of the data. Comparing with a
model galaxy spectrum, we obtain a redshift estimate of 
$z_{\mathrm{lens}}=0.44\pm 0.02$.
\keywords{Quasars: individual: HE~0435$-$1223 --
             Quasars: general --
             Gravitational lensing
            }
}

\maketitle

\section{Introduction}

HE~0435$-$1223{} was first discovered in the course of the Hamburg/ESO survey
for bright QSOs \citep{wisotzki*:00:HES3},
and recently found to be a rather spectacular example of a quadruply 
imaged QSO (\citealt{wisotzki*:02:HE0435}; hereafter
Paper~I). The quasar is at a redshift of $z=1.689$ and had a total magnitude
$g = 17.8$ at the epoch of discovery, with evidence for significant variability.
The conspicuous geometry of four blue point sources of similar
brightness arranged with nearly perfect symmetry around a red elliptical 
galaxy left no doubt about the nature of this object as being gravitationally
lensed. However, Paper~I featured only a blended spectrum of all 
components, and a formal verification based on individual spectroscopy 
of each component is still pending. In this paper we present the
first set of spatially resolved spectra of this quadruple QSO, obtained
by integral field spectrophotometry.

\begin{figure*}[t]
\setlength{\unitlength}{1mm}
\begin{picture}(120,95)
\put(0,62){\includegraphics[width=28mm,bb=98 98 302 302]{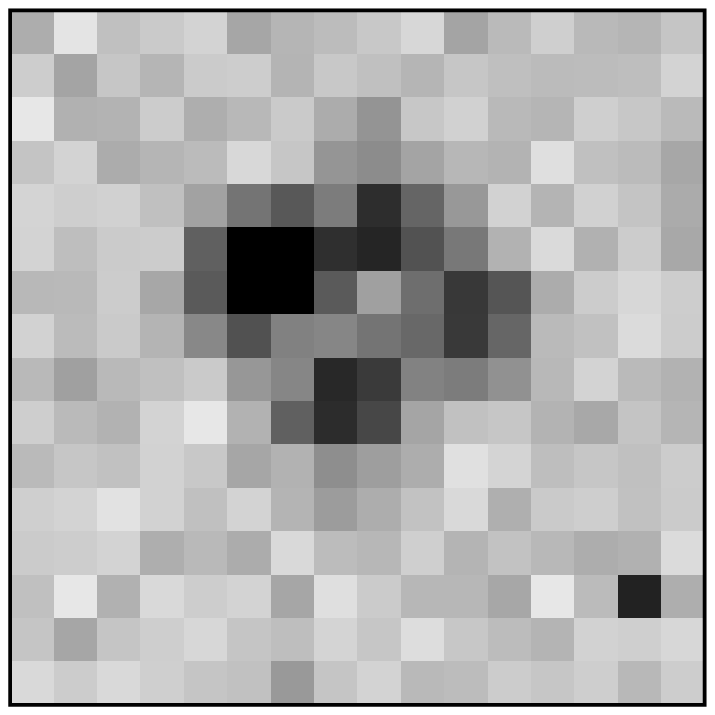}}
\put(1,91){$\lambda = 4278$\,\AA}
\put(30,62){\includegraphics[width=28mm,bb=98 98 302 302]{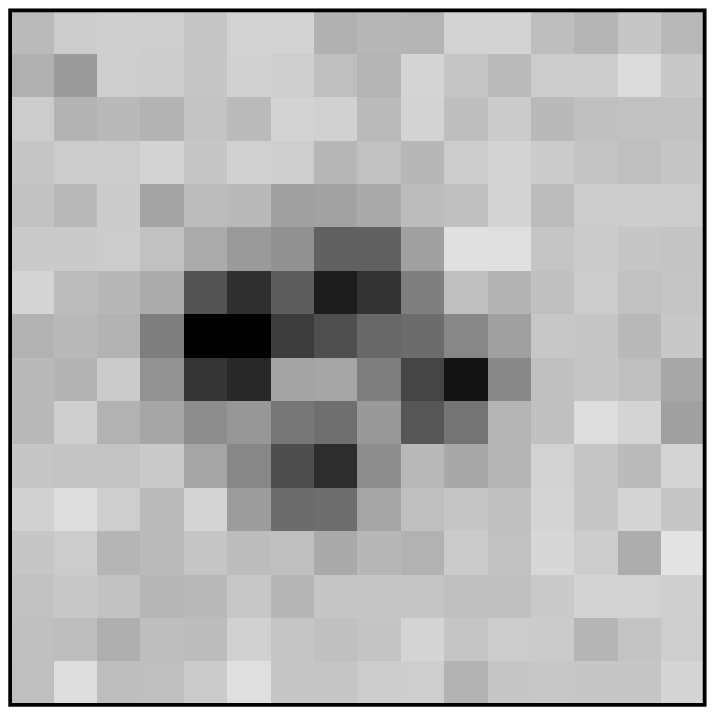}}
\put(31,91){$\lambda = 5601$\,\AA}
\put(60,62){\includegraphics[width=28mm,bb=98 98 302 302]{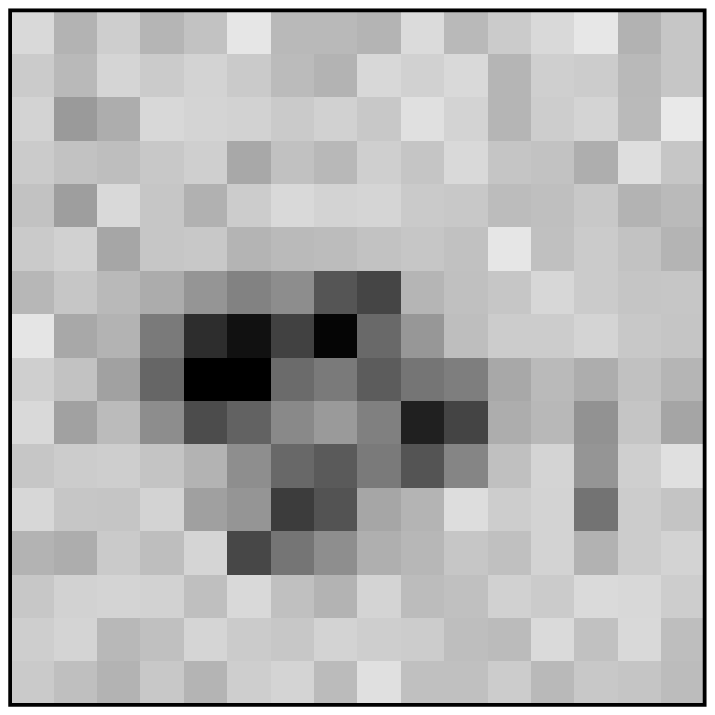}}
\put(61,91){$\lambda = 6924$\,\AA}
\put(90,62){\includegraphics[width=28mm,bb=98 98 302 302]{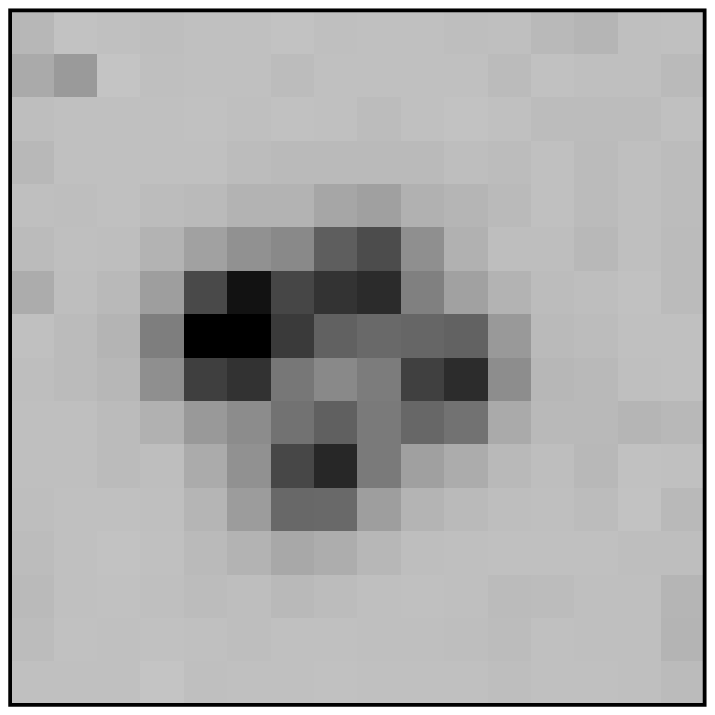}}
\put(91,91){5000--6000\,\AA}
\put(0,31){\includegraphics[width=28mm,bb=98 98 302 302]{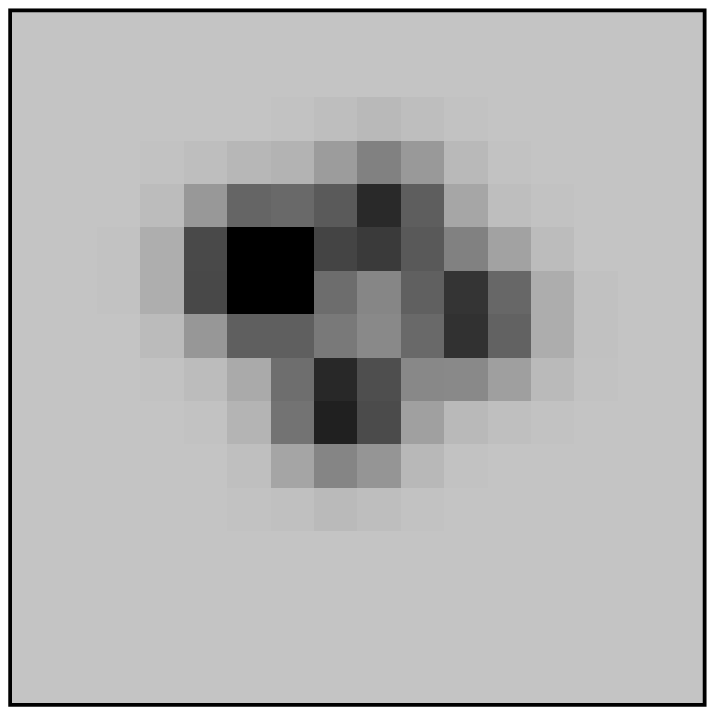}}
\put(30,31){\includegraphics[width=28mm,bb=98 98 302 302]{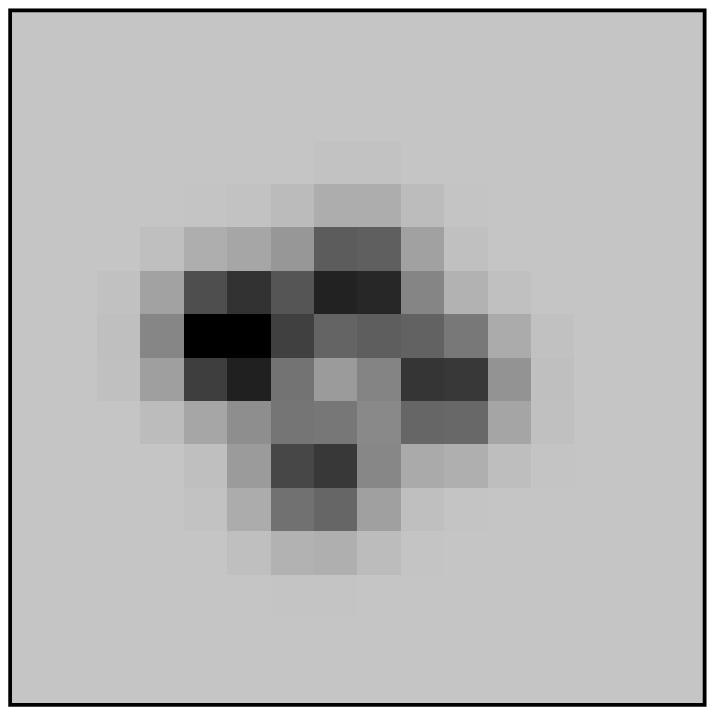}}
\put(60,31){\includegraphics[width=28mm,bb=98 98 302 302]{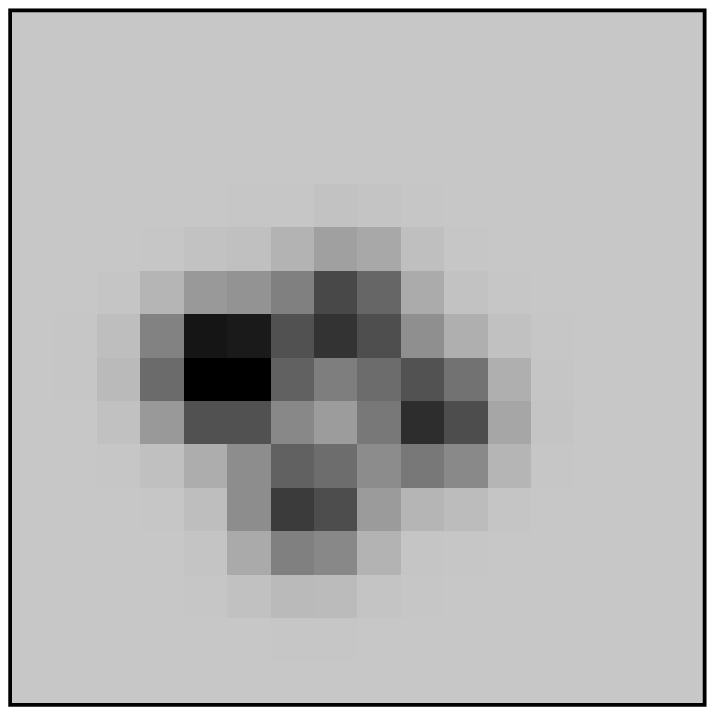}}
\put(90,31){\includegraphics[width=28mm,bb=98 98 302 302]{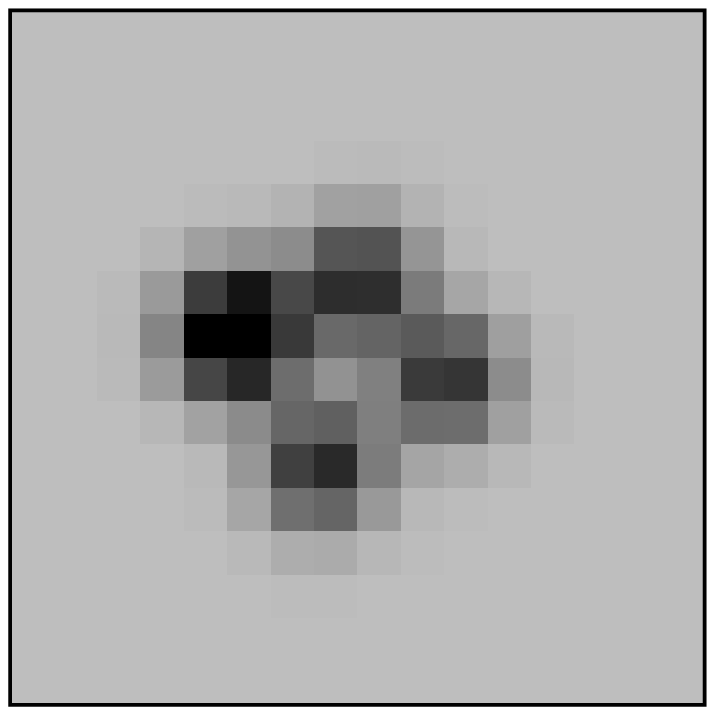}}
\put(0,0){\includegraphics[width=28mm,bb=98 98 302 302]{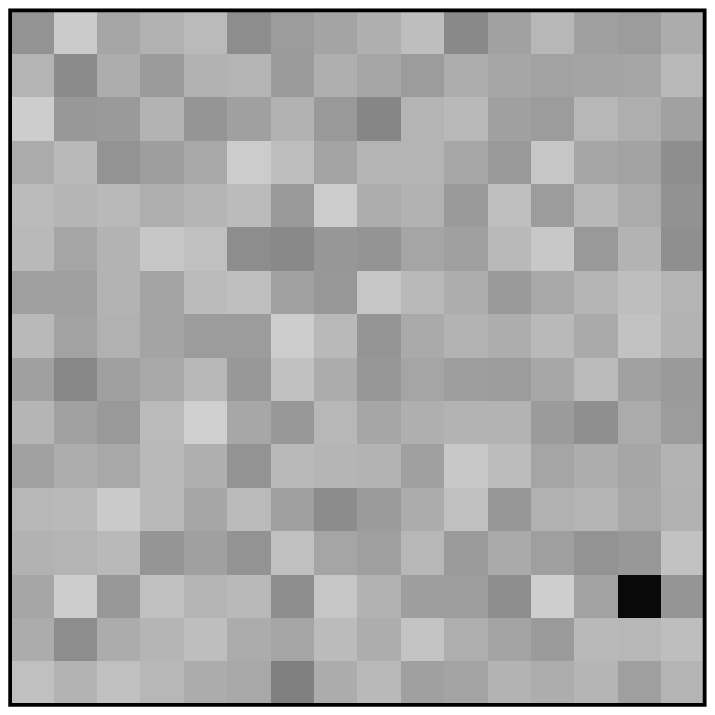}}
\put(30,0){\includegraphics[width=28mm,bb=98 98 302 302]{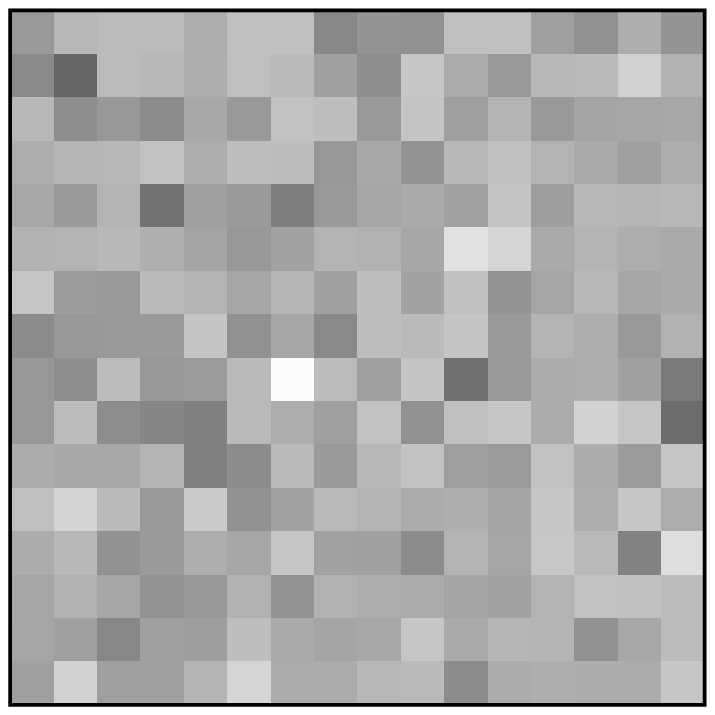}}
\put(60,0){\includegraphics[width=28mm,bb=98 98 302 302]{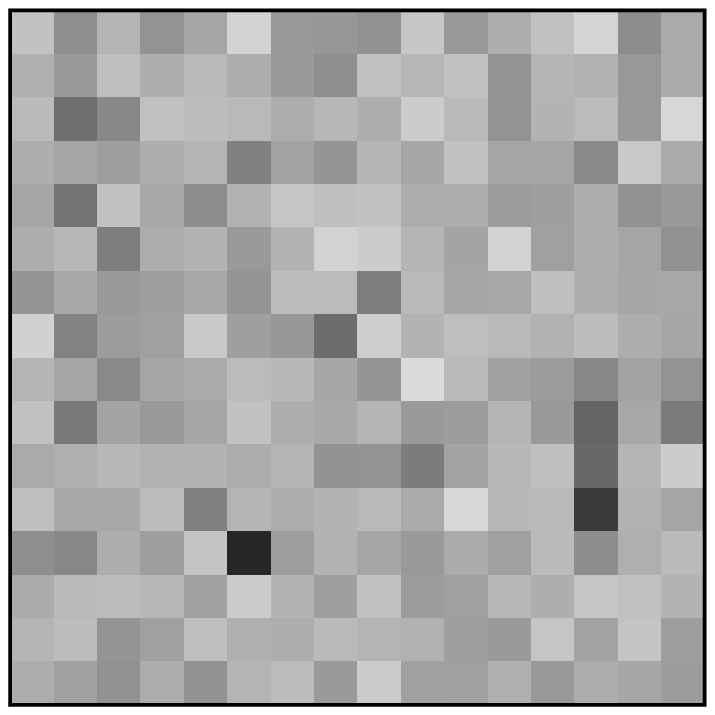}}
\put(90,0){\includegraphics[width=28mm,bb=98 98 302 302]{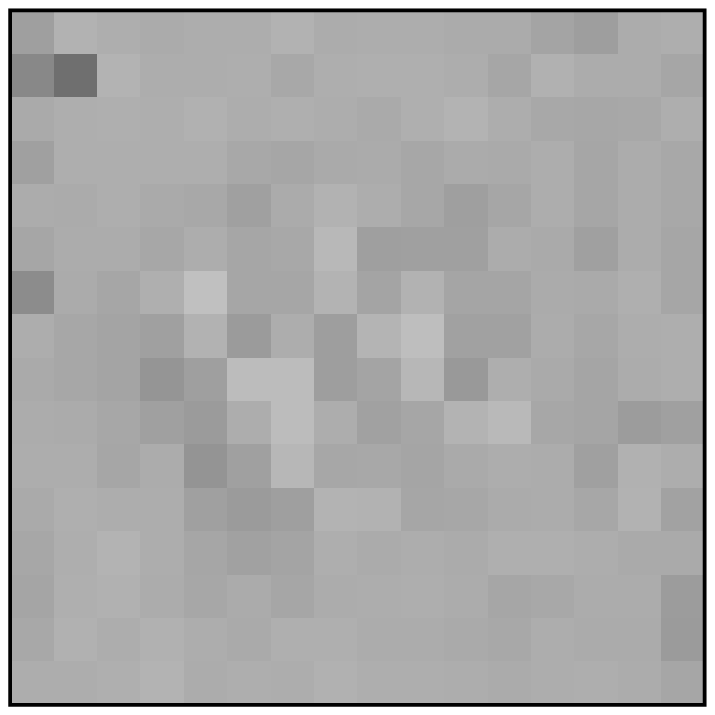}}
\end{picture}
\hspace*{\fill}
\parbox[b]{55mm}{
\caption[]{Top row: Example PMAS data of HE~0435$-$1223, for different wavebands.
The first three images are quasi-monochromatic (3.3\,\AA\ bandwidth),
the right-hand image is averaged over a wide passband. Each image measures
$8''\times 8''$m and each spatial pixel is $0\farcs 5 \times 0\farcs 5$. 
The orientation is standard
(North is to the top, East is to the left).
The four QSO components are clearly resolved. Notice the strong 
shift of image centroids as a function of wavelength (cf.\ Fig.\
\ref{fig:cen}), caused by differential atmospheric refraction.
Middle row: Corresponding model images consisting of four superposed
Gaussians (for details see text).
Bottom row: Residuals after subtracting the four point sources.
}}
\label{fig:ima}
\end{figure*}

\section{Observations and data reduction}    \label{sec:obs}

We targeted HE~0435$-$1223{} during the first regular (i.e.\ non-comissioning)
observing run of the \emph{Potsdam Multi-Aperture Spectrophotometer} PMAS,
mounted at the Calar-Alto 3.5~m telescope for several nights between 
2--7 September, 2002. Details of this new instrument are given elsewhere
\citep{roth*:00:PMAS};
here we only summarise its capabilities. PMAS currently consists of a 
$16\times 16$ elements microlens array (an upgrade to $32\times 32$ elements
is planned) coupled by optical fibres 
to a purpose-designed spectrograph. The image scale is $0\farcs 5$
per spatial pixel (\emph{`spaxel'}), thus the total field of view is 
$8''\times 8''$. We used the V300 grating,
giving a spectral resolution of $\sim 6$\,\AA\ FWHM and a spectral
range of $\sim 3300$\,\AA , from 3950\, to 7250\,\AA\ in our preferred
setting. The data were recorded by a SITe $2048\times 4096$ pixels CCD 
detector, read out in $2\times 2$ binned mode. PMAS also features an
independent on-axis imaging camera with a $3\farcm 4 \times 3\farcm 4$ 
field of view and a $1024\times 1024$ pixels detector, 
mainly used for acquisition and guiding (termed the \emph{AG camera} henceforth). 

At the observing date in early September, 
HE~0435$-$1223{} became visible only very shortly before 
morning twilight, and only at airmasses of around 2. 
We revisited the object repeatedly and could record
useful data at the end of three different nights, 
with exposure times of altogether $3\times 1800$\,s and $1\times 1200$\,s.
The external (DIMM) seeing was usually around or slightly below $1''$,
which translated into an effective seeing in our PMAS data of between
$1\farcs 1$ and $1\farcs 3$. Wavelength calibration and tracing
of the 256 fibre spectra were facilitated through exposures using
an internal calibration unit before or after each science spectrum.

Observations of the spectrophotometric standard star Hz~4 
in the morning twilight provided the data needed for flux calibration.
Since the AG camera was equipped with a Johnson $V$ filter, 
we could also exploit the acquisition exposures to check and improve 
the spectrophotometric calibration accuracy. Despite the high
airmass of HE~0435$-$1223{} during the observations, the flux level measurements
of corresponding quasar spectra taken in different nights were
remarkably consistent, with deviations of the order of 1\,\%.
This indicates both photometric observing conditions and a stable system.
Notice that since the PMAS microlens array reimages the exit pupil
of the telescope, there are no geometrical losses due to incomplete 
filling of the focal plane (as with some other integral-field instruments).
PMAS is thus certainly suitable for accurate spectrophotometry, 
and we shall use this capability below.

The data were reduced using our own IDL-based software package P3d
\citep{becker:02:D}. The reduction consists of 
standard steps such as debiasing and flatfielding using 
twilight exposures, and dedicated routines such as tracing
and extracting the spectra of individual fibres and reassembling
the data in form of a three-dimensional data cube. Wavelength
calibration and rebinning to a constant spectral increment 
is also part of the reduction procedure. The top row of 
Fig.\ \ref{fig:ima} shows as examples some quasi-monochromatic and 
synthetic broad-band images, giving a good representation of the 
image quality, in particular S/N level and angular resolution.
In fact, the S/N measured in our PMAS data is fully consistent with 
expectations from pure shot noise.

\section{Deblending of the QSO components}

\subsection{Procedure} \label{sec:fits}

\begin{figure}[tb]
\includegraphics*[width=83mm]{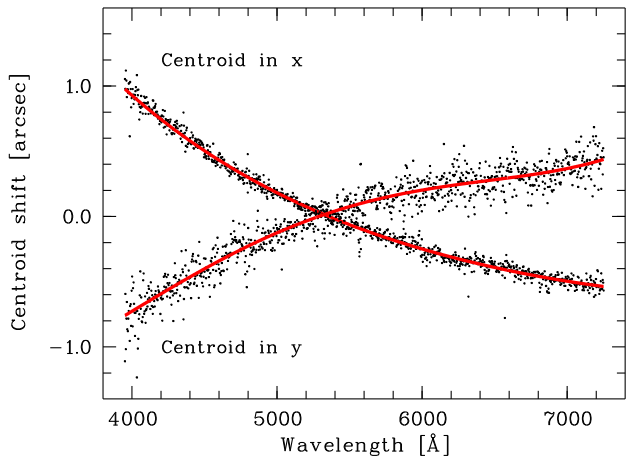}
\caption[]{Centroid displacement as a function of wavelength,
mainly due to differential atmospheric refraction.
Each point corresponds to one quasi-monochromatic spectral
layer. In the later fitting passes, the measured centroids 
are replaced by the polynomial approximation 
plotted by the solid lines.}
\label{fig:cen}
\end{figure}

\begin{figure}[tb]
\includegraphics*[width=83mm]{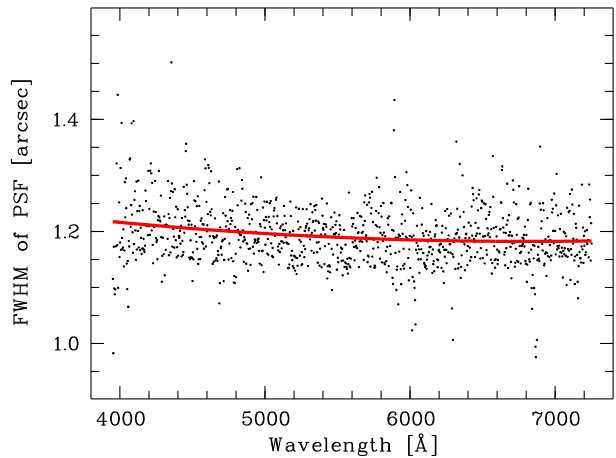}
\caption[]{Same as Fig.\ \ref{fig:cen}, but for the 
full width at half maximum (average of major and minor axis)
of the Gaussian PSF.}
\label{fig:fwhm}
\end{figure}

Starting with the data cubes produced on output by the 
P3d package, we developed a simple but powerful scheme to extract
all four QSO spectra simultaneously. The following 
considerations guided our strategy:

\begin{itemize}
\item Because of the small field of view, the point-spread
function was not known a-priori, but had to be determined
within the QSO data.%
 \footnote{In the future it will be possible to use the
 guide star data recorded by the AG camera during a PMAS
 exposure to establish a simultaneous PSF reference, 
 but this mode was not yet operational at the time of observing.}
Non-simultaneous observations of PSF calibrator stars were
not considered an option, because of the significant temporal
PSF variability at the site.
\item Due to differential atmospheric refraction, the source centroids
vary with wavelength relative to the grid of $16\times 16$ spatial
pixels. This effect is particularly grave for our high-airmass data, 
leading to a shift of more than four pixels (i.e.\ $> 2''$) 
over the spectral range covered (see Figs.\ \ref{fig:ima} and
\ref{fig:cen}).
\item Over the substantial spectral range covered, the width of the 
PSF can not assumed to be constant. This effect turned out
to be less relevant than expected in our data, cf.\ Fig.\ \ref{fig:fwhm},
but significantly affected other data taken in the same campaign.
\end{itemize}
Our adopted algorithm is conceptually similar, albeit somewhat 
more complex, to procedures already used 
by us and others for the simultaneous extraction of 
double QSO spectra in long-slit data 
(e.g., \citealt{smette*:95:GLC}).

In essence, the routine performs a multi-component fit
of $n$ two-dimensional Gaussians plus a spatially constant
background value to the data in each quasi-monochromatic image,
using a modified Levenberg-Marquard minimisation scheme
\citep{press*:92:NR}.
In the case of HE~0435$-$1223{}, we adopted $n=4$ for the four QSO components.
At the chosen spectral range, it was not strictly necessary
to include also the lensing galaxy; as discussed in 
Sect.\ \ref{sec:lens} below, we have marginally detected the
galaxy, although on a level not much above the PSF mismatch 
residuals.

On output, the procedure provides the fitted parameters and 
their standard errors in a table.
Each Gaussian is basically characterised by six parameters 
per spectral element: centroids in $x$ and $y$, FWHM along
major and minor axis, position angle, and amplitude. By including
some prior knowledge we can reduce the number of free parameters:
First, the values of FWHM are assumed to be the same for all 
four point sources. The process is then iterated: 
In an initial pass, all parameters are fitted freely. 
In subsequent passes, the centroids, FWHM values and position angles 
are replaced by smoothly varying polynomial functions of $\lambda$,
as exemplified by Figs.\ \ref{fig:cen} and \ref{fig:fwhm}. 
Once fitted, these parameters are held fixed and do not participate 
in the fitting process, 
leaving only the 4 amplitudes to be fitted in the final pass. 
Model and residual data cubes are constructed along for inspection
(see Fig.\ \ref{fig:ima}).
The procedure turned out to be robust and efficient, simultaneously 
providing deblended spectra and a reasonable PSF, as
demonstrated by the small residuals in the bottom row 
of Fig.\ \ref{fig:ima}.

After the final pass, the product of fitted amplitudes and
(FWHM)$^2$ gave the extracted number of counts 
per spectral element for each QSO component. A similar 
extraction performed for the spectrophotometric standard star
Hz~4 provided a formal flux calibration. We note that 
the efficiency of PMAS at 4000\,\AA\ is still 30\,\% of 
its peak value at 6000--7000\,\AA , which is remarkably
high for an integral field spectrograph.

\begin{figure*}[t]
\includegraphics*[width=180mm]{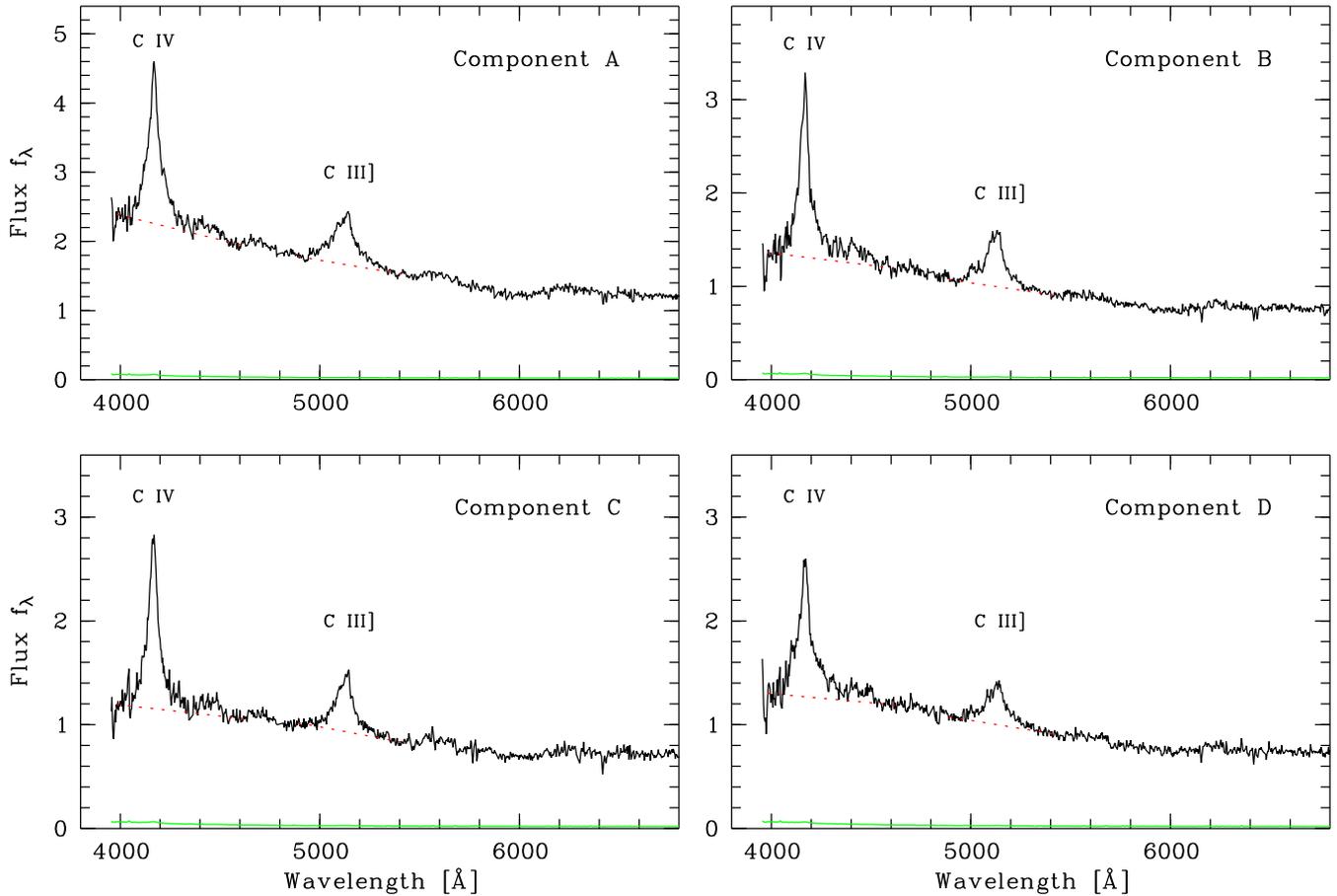}
\caption[]{Extracted and coadded spectra of the four QSO components.
The thin lines near zero represent the 1$\sigma$ error arrays. 
The adopted local continuum levels, used to derive emission 
line properties, are indicated by the dashed line segments.}
\label{fig:spec1}
\end{figure*}

\subsection{Results}

Visual inspection of the images confirms that the four
components are well separated, with little or no
significant mutual contamination (see Fig.\ \ref{fig:ima}). 
The extracted spectra from the four exposures of HE~0435$-$1223{}, 
obtained during three different nights, were combined using
an inverse variance weighting scheme, which was also used
to eliminate bad data values contaminated by cosmics or other
spurious signals. Figure \ref{fig:spec1} shows the four 
final spectra together with the associated error vectors. 
The S/N in the continuum around $\lambda = 5000$\,\AA\ 
is about 60 for component A and $\sim 40$ for the three 
fainter components B, C, and D.

Apparently, all four components show very similar and typical 
quasar spectra. While this is not really a surprise given the
configuration of the system and the colours measured in Paper~1, 
it is nevertheless the first time that individual spectra of 
the components of HE~0435$-$1223{} are published.
The redshifts derived from the \ion{C}{iv}
and \ion{C}{iii}] lines are identical to within the measurement
accuracy: $z = 1.6895 \pm 0.0005$; the redshift scatter  
between the components is $\Delta z \simeq 0.0015$ (rms),
corresponding to less than one spectral pixel.

\section{Spectrophotometric Analysis}

\subsection{Broad-band fluxes} \label{sec:bbf}

Our broad-band data from Paper~1 were taken with
the Magellan telescope in SDSS filters $g$, $r$, and $i$ 
at two epochs, 14~Dec~2001 and 12~Feb~2002.
In order to facilitate a straightforward comparison 
between these and our new PMAS data,
we list synthetic broad-band magnitudes in Table \ref{tab:mag}, 
obtained by integrating over the $g$ and $r$ filter curves 
(our spectra do not cover the $i$ band). 
At the high S/N of our data, the formal errors of the
synthesised broad-band values are far below 0.1\,\% 
and presumably outweighed by unknown systematics.
The overall flux level in the $g$ band increased by 0.5~mag
between Dec~2001 and Sep~2002. Already in Paper~1 we noted
that HE~0435$-$1223{} was variable, having brightened by almost 0.2~mag 
between Dec~2001 and Feb~2002. We did not detect variability
on timescales of days: The fluxes measured in the three
consecutive observing nights during this run are constant 
to within $\sim 1$\,\%.

The magnitude differences between the components are largely 
consistent with the values quoted in Paper~1, with one
exception: Component D has brightened by 0.07~mag,
from being the faintest component in the Magellan data to 
now at the same level as component B. 
We argue below that this is most likely due to microlensing.

We can also provide the first reliable colour measurements.
The colours listed in Paper~1 were based
on a very crude calibration scheme, {tied} to just one
star with rather inaccurately known intrinsic colours.
We find that the colours for all four components
are basically identical, $(g-r)_{ABCD} = 0.35$ with
an rms scatter of 0.02~mag.
In Paper~1 we found component C to be just slightly 
redder than the mean by 0.03~mag; although this is clearly
at the edge of our measurement accuracy, we find 
that $(g-r)_{C} = 0.38$, and the colour deviation
is exactly reproduced in our new spectra. 
{However, these results do not necessarily imply 
differences in continuum slope, as broad-band colours 
may well be influenced, at this accuracy level, by the 
different emission line equivalent widths discussed in the
next section.}

\subsection{Emission lines}

An important spectrophotometric diagnostic for gravitationally
lensed QSOs is the comparison of emission line properties
between components. While ideally the spectra of lensed quasar 
components should look identical, microlensing and other
contaminating effects may lead to differences. 
In order to derive the emission line properties, we had to
define the underlying continuum. Since we were not interested
in a global continuum level, we defined two \emph{ad hoc} 
continuum windows left and right of each line and connected these
by straight lines (see Fig.\ \ref{fig:spec1}). The windows were
located at [3980\,\AA , 4020\,\AA ] and [4530\,\AA , 4630\,\AA ]
for the \ion{C}{iv} $\lambda$1550 line, and 
at [4890\,\AA , 4950\,\AA ] and [5380\,\AA , 5440\,\AA ]
for the \ion{Al}{iii}/\ion{C}{iii}] $\lambda$1909 blend.
The local continua thus derived are indicated by the dashed
line segments in Fig.\ \ref{fig:spec1}. This simple procedure worked
quite well, although the continuum level at the blue side 
of the \ion{C}{iv} line is poorly constrained due to the
short wavelength cutoff of the spectra. 

We then integrated the continuum-subtracted spectra within 
narrow windows centred on the emission line cores. The selection
of these windows was to some extent arbitrary and mainly guided
by the wish to avoid the low S/N wings which are most strongly
affected by errors in the continuum estimate. The finally adopted
integration windows were [4020\,\AA , 4300\,\AA ] for \ion{C}{iv} and 
[5075\,\AA , 5180\,\AA ] for \ion{C}{iii}]. Table \ref{tab:lines}
lists the results for all four components. 
We do not quote formal statistical errors because, as before, 
these are very small, typically $\sim 1$\,\% for the line fluxes, 
and almost certainly dominated by unknown systematical errors.

With high significance, the equivalent widths of the lines
are not identical in the four components. This is already
apparent from visual inspection of the spectra in 
Fig.\ \ref{fig:spec1}. In other words: The flux ratios
between the QSO components are different in the continuum
and in the broad emission lines. We discuss this phenomenon
in the next sections.

\begin{figure*}[t]
\includegraphics*[width=120mm]{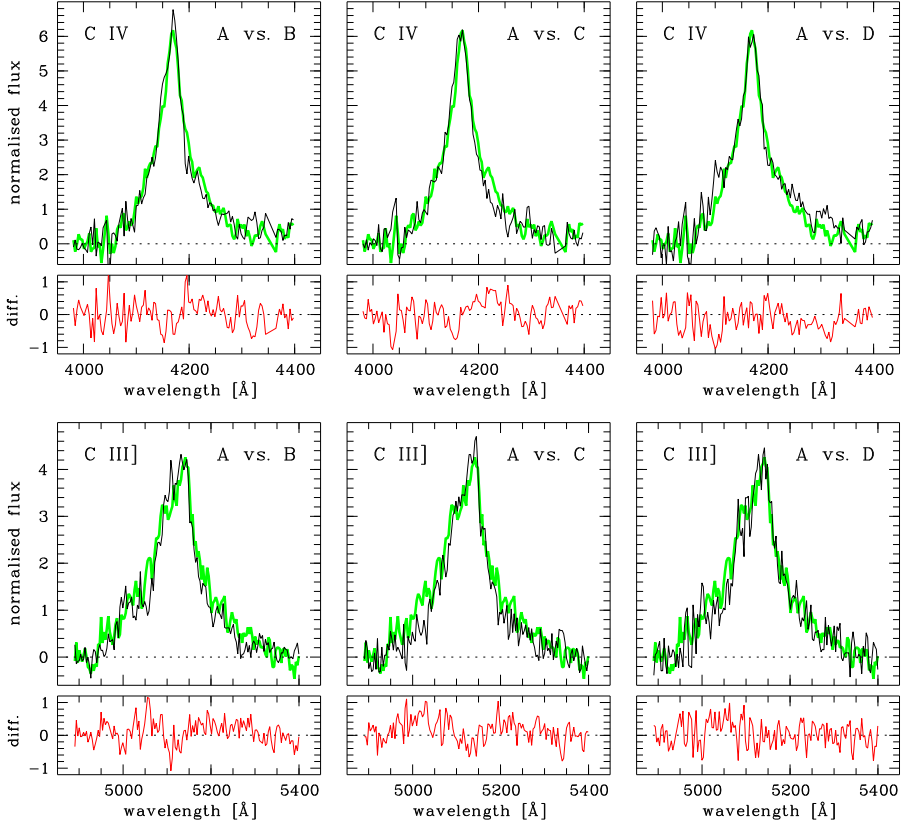}
\hspace*{\fill}
\parbox[b]{55mm}{
\caption[]{Pairwise comparison of normalised emission 
line profiles
(upper row: \ion{C}{iv}; lower row: \ion{C}{iii}]).
The light thick lines represent component A,
the dark thin lines denote components B, C, and D, respectively.
Below each panel we show also the difference between the profiles. }
\label{fig:lines}}
\end{figure*}

Finally, we also compared the emission line profiles. To this
purpose we rescaled all continuum-subtracted lines 
to a common line integral and superimposed them pairwise. 
This is documented in Fig.\ \ref{fig:lines}, 
where the normalised \ion{C}{iv} and \ion{C}{iii}] lines 
of each of the components B, C, and D are plotted over A. 
As expected for a gravitationally lensed QSO, the individual
line profiles are very similar and consistent with the
assumption that they originate in the same object.

\begin{table}
\caption[]{Synthetic broad band magnitudes for HE~0435$-$1223{}, in the
Johnson \emph{V} and in the SDSS $g$ and $r$
bands (Vega magnitudes). Labelling of components as in 
Paper~1, i.e.\ clockwise.}
\label{tab:mag}
\begin{tabular}{lrrrr}
\hline\noalign{\smallskip}
Component & \multicolumn{1}{c}{$V$} & \multicolumn{1}{c}{$g$} 
 & \multicolumn{1}{c}{$r$} & \multicolumn{1}{c}{$g-r$}  \\
\noalign{\smallskip}\hline\noalign{\smallskip}
A         & 18.34 & 18.50 & 18.16 & 0.34 \\
B         & 18.88 & 19.03 & 18.70 & 0.33 \\
C         & 18.96 & 19.15 & 18.77 & 0.38 \\
D         & 18.90 & 19.07 & 18.72 & 0.35 \\
A+B+C+D   & 17.23 & 17.40 & 17.05 & 0.35 \\
      \noalign{\smallskip}\hline
  \end{tabular}
\end{table}

\begin{table}
\caption[]{Emission line fluxes and rest-frame 
equivalent widths for the four components. 
Fluxes are expressed in 
$10^{-16}\:\mathrm{erg}\:\mathrm{s}^{-1}\:\mathrm{cm}^{-1}$,
equivalent widths in \AA .}
\label{tab:lines}
\begin{tabular}{lrr@{\hspace{3em}}rr}
\hline\noalign{\smallskip}
 & \multicolumn{2}{c}{\ion{C}{iv} $\lambda$1550 \hspace*{1.5em}} & 
            \multicolumn{2}{c}{\ion{C}{iii}] $\lambda$1909} \\
\noalign{\smallskip}
Component & \multicolumn{1}{c}{$f$} & \multicolumn{1}{c}{$W_0$\hspace*{2em}}   
 & \multicolumn{1}{c}{$f$} & \multicolumn{1}{c}{$W_0$} \\
\noalign{\smallskip}\hline\noalign{\smallskip}
A & 120.1 & 19.2  & 60.3 & 13.5  \\
B &  90.4 & 24.2  & 46.3 & 17.2  \\
C &  86.0 & 26.8  & 42.3 & 16.9  \\
D &  66.2 & 18.3  & 31.7 & 11.8  \\
\noalign{\smallskip}\hline\noalign{\smallskip}
\end{tabular}
\end{table}

\section{Spectral differences: Evidence for Microlensing}

In the previous section we documented that despite the 
certain nature of HE~0435$-$1223{} as a lensed QSO, the spectra of
individual components are not strictly identical. Several
effects can be invoked to explain spectral differences between
lensed images: Foreground extinction, intrinsic variability, 
or microlensing. We discuss each of these in turn.

Extinction is probably not important in this object.
The nearly identical (and typically quasar-like) colours 
of the four components imply that there is no significant 
differential extinction between these lines of sight. 
Although it is conceivable that all lines of sight suffer 
from the same amount of extinction, this would be 
an extreme coincidence if the dust column density were high.
Furthermore, the lensing galaxy is a large elliptical (cf.\ Paper~1),
not expected to contain large quantities of 
smoothly distributed dust in its outskirts. 
And of course the different equivalent widths of the emission
lines cannot be explained by dust extinction at all, so there
has to be at least one additional effect.

Intrinsic flux variability is certainly an issue in HE~0435$-$1223{},
as already discussed in Sect.\ \ref{sec:bbf}.
However, it is unlikely that variability is the explanation 
for the differences. 
From the lensing geometry, the light travel time delay between 
the images is predicted to be of the order of a few days (Paper~1). 
This means that very rapid and drastic intrinsic spectral changes 
would be needed in order to create differences between the images;
we see no evidence at all for such violent variability.
Even if there were such changes, escaping our attention, 
these would have to appear in each of the components consecutively 
within days. The fact that the spectra taken at three different 
nights look identical within the error bars (i.e.\ with an uncertainty
of no more than a few per cent) speaks against significant 
short-time spectral variability.

Gravitational microlensing remains as the most
straightforward explanation for the observed spectral differences
between the four images.
Within that scenario, microlensing affects the continuum and not
the emission lines because of the sizes involved: The broad-line region
in quasars is believed to be much larger than the typical cross-section
of stellar mass lenses, and any magnification patterns are
thus completely smeared out. On the other hand, the continuum-emitting 
region is orders of magnitude smaller than the BLR and certainly prone 
to microlensing (de)magnification, as most dramatically seen in the 
`Einstein Cross' (e.g., \citealt{wozniak*:00:HCC}). Spectroscopy 
of that quasar \citep{lewis*:98:MISV} showed 
also differences in the emission line equivalent widths
very similar to the case discussed here. The most extreme 
known case of selective continuum amplification by
microlensing is the `Double Hamburger' HE~1104$-$1805
\citep{wisotzki*:93:HE1104}, where there is even a strong 
chromatic effect, i.e.\ a wavelength-dependent amplification 
of the continuum (the spectrum gets bluer because the continuum region 
is smaller at shorter than at longer wavelengths and hence is more
magnified; cf.\ \citealt{wamb+pacz:91:CV}).

For all these objects as well as for HE~0435$-$1223{}, the emission line 
\emph{profiles} are identical in all components. The spectral
differences can then be interpreted as being entirely due to 
additional continuum contributions on top of the intrinsic
`macrolensed' spectra. Since microlensing is time-variable
as a result of non-negligible effective transversal motions,
this additional contribution is expected to vary. 
Interestingly, it is component D for which we find evidence
for a relative brightening between Dec~2001 and Sep~2002; 
that component has the lowest equivalent widths and thus,
in the microlensing picture, the highest continuum amplification.
We conclude that microlensing is the only plausible explanation
for the observed differences in line-to-continuum ratios. 
On the other hand, there is currently no indication of significant
differential \emph{chromatic} continuum amplification, since all 
components have so similar colours.

{
At this point it is useful to reconsider briefly the question 
whether the emission lines might also be affected by microlensing. 
The broad-line region in quasars (BLR) is generally held to be 
much bigger than the continuum region; for a quasar of the intrinsic 
(de-lensed) luminosity of HE~0435$-$1223{}, a BLR radius of the order of 100\,light-days
can be expected (Kaspi et al.\ \citep{kaspi*:00:RM}), two orders of
magnitude larger than the Einstein radius of a solar-mass star
at $z \simeq 0.4$. Any microlensing would therefore act on a small 
region of the BLR only, possibly causing some variations of the line 
profile as an observable signature, but certainly no significant 
overall amplification (cf.\ also \citealt{schn+wamb:90}).
Even that effect cannot be strong in HE~0435$-$1223{}, since the 
line profiles are so similar (cf.\ Fig.\ \ref{fig:lines}).
We therefore assume in the following that a clean distinction between 
the two domains exists: While the continuum source is very likely 
significantly amplified by microlensing, the emission lines remain 
more or less unaffected.
}

\begin{table}
\caption[]{Observed and predicted flux ratios between the four components: 
(1) \ion{C}{iv} and \ion{C}{iii}] emission lines; 
(2) continuum ratios at $\lambda = 4580$~\AA;
(3) broad-band measurement from Paper~1; (4) flux ratios predicted from
the model in Paper~1.}
\label{tab:ratios}
\begin{tabular}{lrrrr}
\hline\noalign{\smallskip}
Ratio & \multicolumn{1}{c}{lines} & \multicolumn{1}{c}{continuum} 
      & \multicolumn{1}{c}{$g$ band} & \multicolumn{1}{c}{\textit{Model}} \\
\noalign{\smallskip}\hline\noalign{\smallskip}
B/A &  0.77 & 0.62 & 0.60 & \textit{1.11} \\
C/A &  0.71 & 0.54 & 0.57 & \textit{1.04} \\
D/A &  0.56 & 0.61 & 0.55 & \textit{0.68} \\[0.5ex]
C/B &  0.93 & 0.87 & 0.95 & \textit{0.94} \\
D/B &  0.73 & 0.98 & 0.93 & \textit{0.61} \\[0.5ex]
D/C &  0.79 & 1.12 & 0.97 & \textit{0.65} \\
\noalign{\smallskip}\hline\noalign{\smallskip}
\end{tabular}
\end{table}

\section{Flux ratios}

One interesting consequence of the microlensing interpretation
is the expectation that the emission line fluxes 
should mirror the `true' flux ratios of the four components, 
i.e.\ those given by macrolensing of the smooth galaxy potential alone. 
This is important for the construction of mass models for the
lens: While flux ratios from broad-band photometry are
notoriously unreliable because of the ever present possibility
of microlensing, we suggest that emission line flux ratios 
can be directly used as additional model constraints. 
In Table \ref{tab:ratios} we provide the pairwise emission
line flux ratios between the components.
As a kind of `sanity check', we also determined the 
corresponding continuum flux ratios, 
listed in the second column of Table \ref{tab:ratios}.
Since the spectral slopes are so similar, these values do
not change significantly over the spectral range covered.

In Paper~1 it was found that a simple model involving a singular
isothermal sphere plus external shear could reproduce the
positions of the QSO components and of the lensing galaxy
well within the error bars. However, the model (which was
based on astrometric constraints only) predicted components
B and C to be substantially brighter, by more than 0.5~mag 
relative to A and D, than observed in broad-band colours
(cf.\ the last two columns in Table \ref{tab:ratios}).
Such disagreement between models and data has been seen
in several other lensed QSOs as well, and two hypotheses
have recently been proposed to account for these anomalies. 
On the one hand, they may indicate the presence of 
substructure in the lensing potential 
(e.g., \citealt{metc+zhao:02:FR}); 
or they could be due to microlensing at large optical depths,
especially by suppressing the flux of saddle point images
\citep{sche+wamb:02:QMHM}. These two explanations differ
significantly in their prediction of the flux ratios
of \emph{extended} source components.
While the flux ratios should not depend strongly on source 
size in the case of a lens with substructure, 
a necessary consequence of the microlensing 
hypothesis is that the flux ratios should approach the 
`macrolens' values as soon as the source is too extended 
to be affected by the microlens caustic network. 
This implies that if the size of the QSO broad-line region 
is already in this domain, emission-line flux ratios should be 
much closer to the model values if the model is realistic.
{
There is at least one known quadruple QSO where this seems to be 
the case, namely B~1422$+$231 (cf.\ \citealt{impey*:96:B1422}).
}

This test can be performed with our spectrophotometric 
measurements of HE~0435$-$1223{}. Table \ref{tab:ratios} shows that
the agreement between measured flux ratios and the model predictions 
is indeed much better if one uses the emission line values 
than with either broad band or continuum data. 
Components B and C are brighter in the emission lines than 
in the continuum, relative to A and D, while the ratios A/D and B/C 
do not change significantly. These changes are precisely in the 
direction approaching the model, just as predicted in the microlensing
scenario by Schechter \& Wambsganss. 

However, the agreement is anything but good, and especially
component A is still much brighter than is expected. 
As this could be due to shortcomings of the simple model which did not
even take flux ratio constraints into account, we attempted
to compute a new model. We used C.~Keeton's software tool
\texttt{lensmodel} \citep{keeton:01:CM} 
to explore the consequences of adding measured flux ratios 
as additional constraints. For consistency, we stayed within
the singular isothermal sphere plus external shear model
(although we also tried some elliptical models). To our
surprise, it was extremely difficult to make the predicted 
flux ratios approach the measured ones; the predictions
tended to stay close to the Paper~1 values, which were based 
on the astrometry {of the four QSO components} alone
{(the position of the lensing galaxy was not considered
to be sufficiently constrained, and therefore it was always 
left to float, as in Paper~1)}.
This would only change when the flux ratio constraints were taken 
to be \emph{much} tighter than the astrometric constraints.
For example, we could roughly fit the flux ratios when allowing for
several tenths of arc seconds of positional shifts,
corresponding to a formal deviation of the order of more than
$\sim 50\sigma$, which is totally unacceptable given the excellent 
astrometric quality of the Magellan data.
Thus, even using the emission line measurements we failed to build
a consistent model for the lensing potential that correctly
reproduces both positions and flux ratios. 
But we recognise that a more sophisticated model of the lens
mass distribution might well prove this statement premature. 
We have deliberately limited ourselves to explore 
only a narrow range of plausible configurations 
and now invite the experts in the field to follow up on us. 
Until then, we conclude that our measurements give some support 
to the microlensing hypothesis by \citet{sche+wamb:02:QMHM}, 
but that no wholly satisfying explanation for the flux ratios 
in HE~0435$-$1223{} can yet be given.

\begin{figure}[t]
\setlength{\unitlength}{1mm}
\begin{picture}(88,45)
\put(0,0){\includegraphics[width=42mm,bb=98 98 402 402,angle=90]{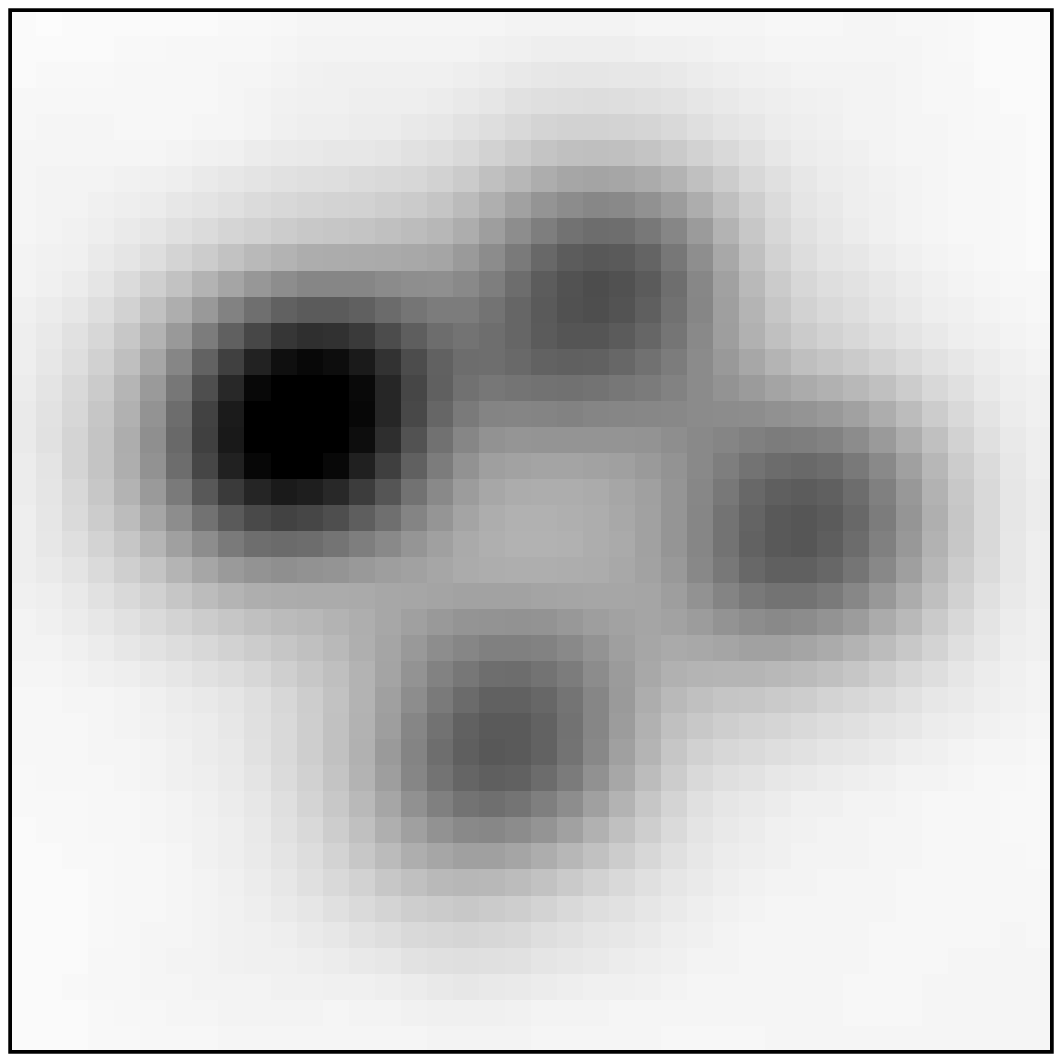}}
\put(46,0){\includegraphics[width=42mm,bb=98 98 402 402,angle=90]{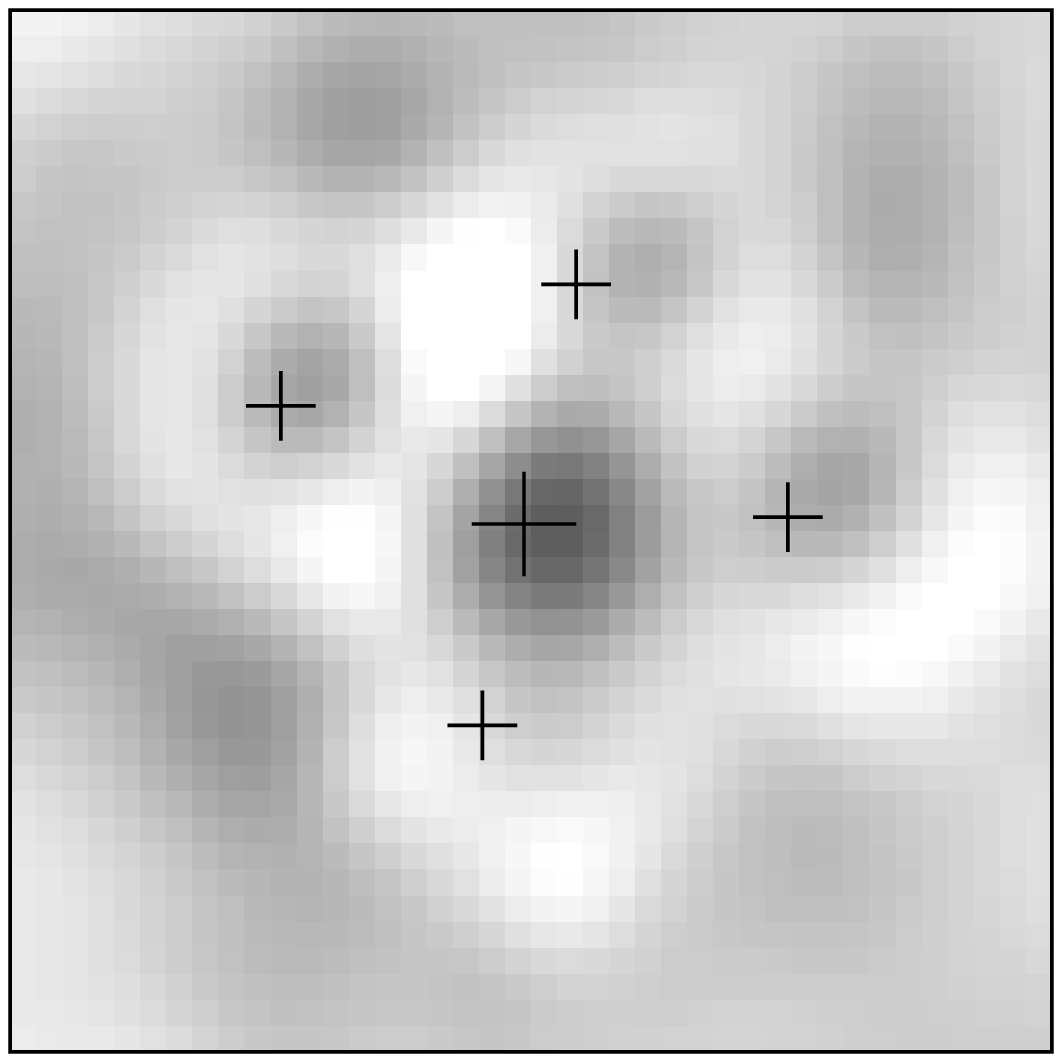}}
\end{picture}
\caption[]{Left panel: $4\times 4$ supersampled image of HE~0435$-$1223{}, 
coadded over a simulated 5500--7200\,\AA\ passband (see text for details).
Right panel: Residual image after PSF subtraction, supersampled and coadded
in the same way. Despite the significant PSF residuals, the lensing galaxy 
is clearly detected. The positions of the four QSO components and the
lensing galaxy (from Paper~1) are marked by crosses.}
\label{fig:super}
\end{figure}

\begin{figure}[t]
\setlength{\unitlength}{1mm}
\includegraphics*[width=\hsize]{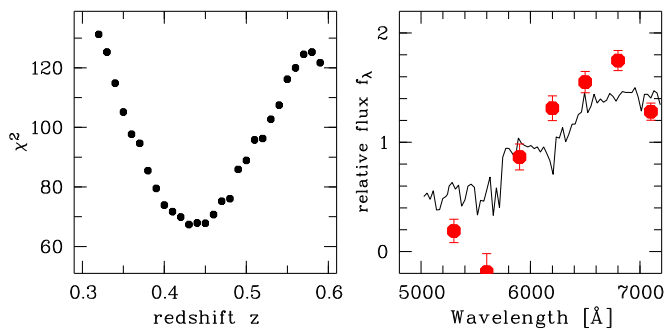}
\caption[]{Redshift estimation for the lensing galaxy. Left panel:
$\chi^2$ vs.\ redshift for model fit of 10~Gyr SSP to the 
binned residual spectrum at position of the lensing galaxy.
Right panel: Residual spectrum with best-fit SSP model 
spectrum superimposed. For details see text.}
\label{fig:gal}
\end{figure}

\section{The lensing galaxy}  \label{sec:lens}

The Magellan $i$ band image presented in Paper~I shows the lensing
galaxy prominently, close to the geometric centre between 
the four QSO images. A similarly clean detection 
in our new PMAS data was not to be expected, 
{for} two reasons: (i) The much coarser pixel 
grid (0\farcs 5 for PMAS vs.\ 0\farcs 07 for Magellan+MagIC); 
(ii) the inferior effective seeing ($\ga 1\farcs 1$ here
vs.\ 0\farcs 6 for Paper~I). Furthermore, the nonavailability
of a simultaneously obtained PSF reference made a clean PSF
subtraction difficult. We have nevertheless attempted to detect
the lensing galaxy, both in `imaging' and in `spectroscopy' mode.
In our search we made use of some pieces of information 
available from Paper~I. Firstly, we know the location
of the galaxy relative to the QSO point sources. Secondly, we
assume that its redshift is located within $0.3 \la z \la 0.5$,
to be in accordance with the photometric redshift estimate in Paper~I.
{Notice that the narrowing of the range to $0.3 \la z \la 0.4$
in Paper~I using a luminosity-mass argument should be disregarded,
as there was an error in computing the rest-frame absolute magnitude.}

In order to partly overcome the coarse pixel grid limitation,
we have exploited the fact that due to atmospheric
refraction, the QSO centroid gradually shifts as a function of
wavelength. In a way, this can be seen as almost equivalent to 
extensive dithering with respect to a broad-band image.
The image is shifted by several pixels over the entire wavelength range, 
a fortuitous byproduct of the high airmass during observation.
Within a band of just $\sim 1000$~\AA , the displacement
is already a good fraction of a pixel.
Furthermore the number of samples taken is huge, 
roughly 300 for a 1000\,\AA\ bandpass. The information volume
required for substantial subsampling therefore certainly exists.
We have written a small application to coadd monochromatic
data planes into a broad-band image, including a $4\times 4$ 
subsampling and applying the polynomial approximation to the
centroid shifts described in Sect.\ \ref{sec:fits}.

Guided by the redshift prior on the lensing galaxy, we 
selected only the red part of the spectrum, 
$5500\,\mbox{\AA} < \lambda < 7200\,\mbox{\AA}$.
In addition, pieces of the spectrum heavily contaminated by
night sky lines were left out. Both the original
and the residual data cubes were run through this procedure.
The result is shown in Fig.\ \ref{fig:super}. The left panel
shows that the four point sources are now much better defined
than in the original coarse pixel data. In the right panel
we find significant PSF residuals, featuring indications of 
positive central elevations and negative anullus-like patterns
which document the limits of our simple assumption to model 
the PSF as a Gaussian. 
But we also find that the strongest residual is always positive 
and occurs very close to the expected position of the 
lensing galaxy. Relative to the position of component A,
the difference from the value quoted in Paper~1 
is only $\sim 0\farcs 1$ in right ascension and 
$0\farcs 2$ in declination.
Going through the same exercise with a 
$\lambda < 5500\,\mbox{\AA}$ image does not yield 
any discernible signal at that location.
We therefore conclude that the detection is real and significant.

Thus encouraged, we looked whether spectral information,
possibly even the lens redshift could be derived. 
We therefore extracted and coadded the spectra 
of a $2\times 2$ pixel block ($1'' \times 1''$) at the expected
location from each residual data cube. The result was an extremely 
noisy spectrum, with a signal that was on average positive 
above 6000\,\AA\ (S/N of $\sim 2$, for the original 3.3\,\AA\ 
wavelength steps) and zero or slightly negative below that. 
No QSO emission lines were present, indicating that cross-talk
from the QSO point sources was negligible.
Since it was hopeless to detect stellar absorption lines from
the lensing galaxy in such low S/N data, we degraded the spectral
resolution by summing the data into bins of 300\,\AA\ width,
with the intention to use the overall spectral energy distribution
for a redshift estimate. This is documented in Fig.\ \ref{fig:gal}.
We assumed an old stellar population to dominate the SED of the
lens (which has elliptical morphology, cf.\ Paper~I), and
cross-correlated the data with a solar metallicity, age of 10~Gyrs 
model galaxy spectrum from \citep{jimenez*:00:PS},
kindly provided by R. Jimenez. 

The left panel of Fig.\ \ref{fig:gal}
shows that $\chi^2(z)$ has a strong but flat-bottomed minimum 
between $z=0.43$ and $z=0.45$, and a turnover at $z=0.58$ 
towards another local minimum at much higher $z$. 
A lens redshift $z>0.5$ is not compatible
with our prior assumptions; {apart from} its incompatibility with the colour 
measurement of Paper~I, such a redshift would make the lens {extremely}
massive. We therefore settle at a redshift estimate of 
$z_{\mathrm{lens}} = 0.44 \pm 0.02$, for which we show our best fit
model overplotted over the binned residual spectrum.
Note that we obtain consistent results (within the quoted error range)
when analysing the individual exposures taken in different nights. 

The model roughly reproduces the main feature of the data, its 
steep rising towards the red, but otherwise the fit is not good,
and the error bars are probably underestimated. Selecting a younger
stellar population would further degrade the fit quality, since
already now our data are too red compared to the 10~Gyrs model
(which is already the age of the universe at $z = 0.44$).
We conclude that our new lens redshift is a significant improvement 
over the rough colour estimate of Paper~I, but the value is nevertheless 
tentative. A measurement of $z_{\mathrm{lens}}$ involving
the detection of stellar absorption lines in the lensing galaxy
is still indispensable.

\section{Conclusions}

We have presented the first spatially resolved spectra of the
new quadruple QSO HE~0435$-$1223{}. The choice of an integral field
spectrograph, and in particular the design of PMAS, enabled
us to obtain high signal-to-noise data of spectrophotometric
quality for all four quasar images simultaneously. The increasing
availability of such instruments at many major telescopes
opens a new window of opportunity to very efficiently 
observe gravitationally lensed QSOs -- especially quadruple
systems which are very hard to observe with traditional slit
spectroscopy. Yet, two limitations need to be overcome to make
integral field systems fully superior to other spectrographs, 
as far as gravitationally lensed quasars are concerned.

Firstly, resolving close systems requires an accurate knowledge
of the point spread function. This is easily supplied by 
modern imaging cameras and multi-slit spectrographs,
but the field of view of existing integral field units is
generally much too small to simultaneously capture an 
isolated PSF reference star. One possible way out would be
to use a second imaging device in parallel, such as is planned
with the AG camera of PMAS. Another option might be to implement
clever `PSF self-calibration' algorithms, alleviating the
need for an external PSF reference. Our simple approach 
involving just an analytic expression could be no more than
a preparatory study to such techniques.

Secondly, there is usually a {difficult trade-off} between
reaching a maximum field of view and the effective pixel
size in the resulting images. Under excellent seeing conditions,
a critical sampling of the PSF requires pixels no bigger
than 0\farcs 2, while most integral field spectrographs,
including PMAS, have \emph{spaxels} at least twice that size. 
As we have shown in this paper, image enhancement by supersampling 
to a finer pixel grid may be possible as a positive byproduct of 
differential atmospheric dispersion effects.
But of course, the gain thus obtained will always be limited by
the unavoidable convolution with the true pixel size.

With these provisions, a new quality level in the astrophysical 
study of microlensing in QSOs can be reached. Without major losses in 
observing time, simple broad band imaging can be replaced by a
detailed spectrophotometric investigation. In this paper we
have presented two pieces of evidence that microlensing is
important in HE~0435$-$1223{}: spectral differences and variability.
Future monitoring of microlensing-induced spectral variations
in this object, and in similar targets, will constrain the 
structural properties of quasars on angular scales of 
sub-microarcseconds. Integral field spectroscopy is likely
to become a tool of choice for this type of research.

\begin{acknowledgements}
We thank P. Schechter and J. Wambsganss for illuminating 
discussions on the issue of flux ratio anomalies, 
{and an anonymous referee for helpful comments}.
PMAS was partly financed by BMBF/Verbundforschung unter 
053PA414/1 and 05AL9BA1/9.
TB, LC, and AK acknowledge support from the ULTROS project
through grant 05AE2BAA/4, also Verbundforschung. 
SFS acknowledges support through a Euro3D Research Training Network 
on Integral Field Spectroscopy,
funded by the European Commission under contract No. HPRN-CT-2002-00305.
AH is supported by the DFG under grant Wa 1047/6-1 and -3.
KJ, AH and LW acknowledge a DFG travel grant under Wi 1369/12-1.
\end{acknowledgements}

\bibliographystyle{aa}
\bibliography{lutzbib,aipbib,ownpubrefjourn,ownpubproceed,ownpubmisc,ownpubthesis}

\end{document}